\documentclass[a4paper,fleqn,usenatbib]{mnras}

\usepackage[T1]{fontenc}
\usepackage{ae,aecompl}

\usepackage{graphicx}	
\usepackage{amsmath}	
\usepackage{amssymb}	

\usepackage{multirow}
\usepackage{booktabs}

\title[The full evolution of supernova remnants]{The full evolution of supernova remnants in low and high density ambient media}

\author[Jim\'enez et al.]{
Santiago Jim\'enez,\thanks{E-mail: sjimenez@inaoep.mx}
Guillermo Tenorio-Tagle,
Sergiy Silich
\\
Instituto Nacional de Astrof\'isica, \'Optica y Electr\'onica, AP 51, 72000 Puebla, M\'exico\\
}

\date{Accepted XXX. Received YYY; in original form ZZZ}

\pubyear{2019}

\defcitealias{1999Mckee}{TM99}

\begin{document}
\label{firstpage}
\pagerange{\pageref{firstpage}--\pageref{lastpage}}
\maketitle

\begin{abstract}
Supernova explosions and their remnants (SNRs) drive important feedback mechanisms that impact considerably the galaxies that host them. Then, the knowledge of the SNRs evolution is of paramount importance in the understanding of the structure of the interstellar medium (ISM) and the formation and evolution of galaxies. Here we study the evolution of SNRs in homogeneous ambient media from the initial, ejecta-dominated phase, to the final, momentum-dominated stage. The numerical model is based on the Thin-Shell approximation and takes into account the configuration of the ejected gas and radiative cooling. It accurately reproduces well known analytic and numerical results and allows one to study the SNR evolution in ambient media with a wide range of densities $n_{0}$. It is shown that in the high density cases, strong radiative cooling alters noticeably the shock dynamics and inhibits the Sedov-Taylor stage, thus limiting significantly the feedback that SNRs provide to such environments. For $n_{0}>5 \times 10^{5}$ cm$^{-3}$, the reverse shock does not reach the center of the explosion due to the rapid fall of the thermal pressure in the shocked gas caused by strong radiative cooling.

\end{abstract}




\section{Introduction}
Supernova Remnants (SNRs) are powerful sources of mass, momentum and energy. They shape the interstellar medium (ISM) of their host galaxies, determine the evolution of the ISM chemical composition and are sources of cosmic rays, radio and X-ray emission \citep[e.g.][]{mckee1977theory, tenorio2015supernovae, 2015MNRAS.451.2757W, elmegreen2017globular}. It has also been suggested that SNRs are effective dust producers (e.g. \citealt{2001Todini}, \citealt{dust1}, \citealt{Bianchi2007}, \citealt{dust2}, \citealt{Micelotta2016}).\\
\begin{figure*}
	\includegraphics[width=1.55\columnwidth]{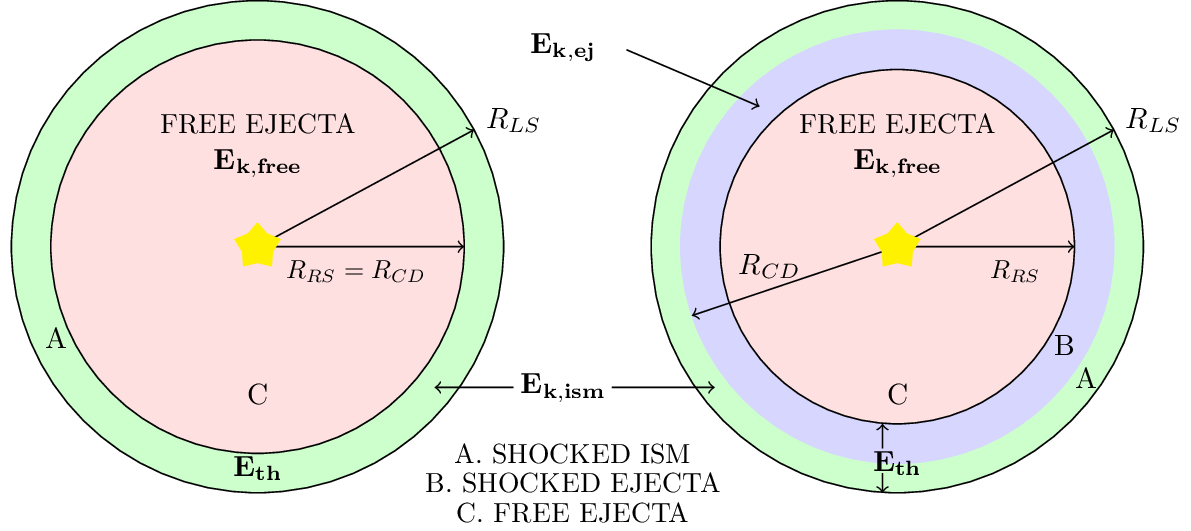}
    \caption{The structure of a SNR. The left panel shows the initial condition at $t=t_{0}$. The right panel presents the structure of the SNR at a later time $t>t_{0}$. See the text for a discussion on the labels of this scheme.}
    \label{fig1}
\end{figure*}
\indent The SNRs undergo several evolutionary stages when the progenitor star explodes in an uniform density media \citep{1977Chevalier}. The first stage, known as the ejecta-dominated (ED) or thermalization phase, begins when the high velocity supernova ejecta collides with the ambient gas and forms a leading shock. The large thermal pressure behind this shock leads to the formation of another, reverse shock, which decelerates and thermalizes the ejected matter. At this stage, the velocity and density structure of the ejecta strongly affects the dynamics of the SNR (e.g. \citealt{draine1993theory}, \citealt{2017Tang}).\\
\indent In low density media, after the reverse shock reaches the center of the explosion and the thermalization process is terminated, the adiabatic Sedov-Taylor (ST) stage begins (e.g \citealt{sedov1993similarity}, \citealt{astrowaves}, \citealt{bisno1}). This stage is described by a self-similar hydrodynamic solution: the shock radius and velocity are given by  power-law functions of time ($R\propto t^{2/5}$, $V\propto t^{-3/5}$ ) and the kinetic ($E_{k}$) and thermal ($E_{th}$) energies are conserved ($E_{k}\approx0.3 E_{0}$ and $E_{th}\approx0.7 E_{0}$, where $E_{0}$ is the explosion energy, see \citealt{sedov1993similarity}). This solution has been widely used as initial condition in many SNR evolution models \citep[e.g.][]{1975MNRAS.172...55F, KimOstriker}, neglecting thus the thermalization phase and assuming that all remnants enter the ST stage. This, as shown here, is not always the case. \\
\indent In homogeneous media, the leading shock slows down with time during the ST stage such that the post-shock temperature reaches values close to the maximum of the cooling function (e.g. \citealt{Raymond1976}, \citealt{Wiersma2009}, \citealt{Schure2009}). Therefore, at late times radiative cooling becomes important. When this occurs, the remnant enters the snowplough (SP) phase  (e.g.  \citealt{CioffiMckee},  \citealt{Blondin1998}, \citealt{mihalas2013foundations}). At this stage, a very thin, cold and dense shell is formed at the outer edge of the SNR. In order to preserve pressure, the density increases in response to the sudden fall of the post-shock temperature. \\
\indent Throughout the course of the SP stage, a SNR loses most of its thermal energy. The remnant then moves for a while in the momentum conserving stage (MCS) and finally merges with the surrounding medium when its expansion velocity becomes comparable to the sound speed in the ambient gas.\\
\indent The evolutionary tracks described above reproduce the physical conditions in many SNRs (e.g \citealt{2000ApJ...528L.109H}, \citealt{slane2000chandra}, \citealt{borkowski2001supernova}, \citealt{Laming2003}). However, the standard theory does not consider the radiative losses of energy at the early ED-phase which, as we show bellow, dominate the SNR evolution when the explosion occurs in a high density medium. Indeed, \cite{terlevich1992starburst} considered a particular case of an ambient gas with density $n_{0}=10^{7}$ cm$^{-3}$, and found that the SNR does not reach the ST stage in such a case.\\
\indent Here we present a model based on the Thin-shell approximation \citep[e.g.][]{astrowaves, silich1992,bisno1} that allows one to follow the full evolution of SNRs, i.e., the thermalization of the SN ejecta, the Sedov-Taylor and the Snowplough stages.  The numerical scheme includes both the initial distribution of density and velocity in the ejecta and radiative cooling of the shocked gas. Our results are tested against well known numerical and analytic results. This model allows us to study the SNRs evolution for a wide range of ambient gas densities ($1$ cm$^{-3}$ $\leq n_{0} \leq$ $10^{7}$ cm$^{-3}$). \\
\indent The paper is organized as follows. In section \ref{sec2}, we introduce the numerical model: the equations of motion and energy conservation and the set of initial conditions. In Section \ref{sec3}, we compare our results with previous numerical and analytic results. Section \ref{sec4} discusses the impact of the ambient gas density on the evolution of SNRs. It is shown that the Sedov-Taylor stage does not occur for SNRs that evolve in densities larger than $5 \times 10^{5}$ cm$^{-3}$ and that scaling relations for the SNRs evolution are not applicable for ambient gas densities above this value. The main differences with the standard evolutionary tracks are addressed including those from high density runs with different metallicities. Finally, Section \ref{sec5} summarizes and discusses our main findings. 
\section{Model Set-up}\label{sec2}
We model the evolution of a SNR in a homogeneous medium with a number density $n_{0}$ from the Ejecta-Dominated to the Snowplough stages. Fig.~\ref{fig1} presents a schematic illustration of the initial condition (left panel) and the resultant SNR structure (right panel), which inside-out presents: the free ejecta, the shocked ejecta and the shocked ambient gas, with kinetic energies $E_{k,free}$, $E_{k,ej}$ and $E_{k,ism}$, respectively. The two outer zones of shocked gas are separated by a contact discontinuity $R_{CD}$. The shocked gas (A and B) loses its thermal energy $E_{th}$ due to radiative cooling. The instabilities of the gas flow are not considered here and therefore no mass traverses the contact discontinuity.  
\subsection{The mass and momentum conservation equations}\label{massConserva}
The evolution of the leading shock is determined by the mass and momentum conservation equations (see \citealt{bisno1} and references therein for a discussion on the Thin-Shell approximation), which in the adiabatic case are:
\begin{equation} \label{eq:1}
\frac{dM_{s1}}{dt}= \rho_{0} U_{LS} 4 \pi R_{LS}^{2},
\end{equation}
\begin{equation}\label{eq:2}
\frac{d}{dt}\left(M_{s1}U_{s1} \right)=4 \pi P R_{LS}^{2},
\end{equation}
\begin{equation}\label{eq:3}
\frac{d R_{LS}}{dt}=U_{LS},
\end{equation}
where:
\begin{equation}\label{eq:3a}
U_{LS}=\frac{\gamma+1}{2}U_{s1}.
\end{equation}
In these equations, $R_{LS}$ and $U_{LS}$ are the leading shock radius and velocity, $M_{s1}$ and $U_{s1}$ are the mass and velocity of the swept-up ambient gas, $\gamma=5/3$ is the specific heats ratio, $\rho_{0}=\mu n_{0}$ is the ambient gas density, $\mu=14/11 m_{H}$ is the mean mass per particle in the neutral gas with 10 hydrogen atoms per helium atom and $P$ is the thermal pressure of the shocked ambient gas. \\
\indent Note that as the swept-up gas cools down, the remnant enters to the SP stage, and in such case equation (\ref{eq:3a}) becomes $U_{LS}=U_{S1}$. The transition to this phase occurs at the \textit{thin shell-formation time} $t_{sf}$, which is the time when the swept-up gas begins to collapse into a cold, dense shell (e.g. \citealt{CioffiMckee}). If an element of gas is shocked at time $t$, it cools at:
\begin{equation}
t_{c}=t+\Delta t_{cool} \left(t \right), 
\end{equation}
where $\Delta t_{cool} \left(t \right) $ is the gas cooling time (e.g. \citealt{Petruk2006, KimOstriker}):
\begin{equation}
\Delta t_{cool} \left(t \right) =\frac{1}{\gamma+1}\frac{k_{B}T_{LS}}{n_{0}\Lambda\left(T_{LS}\right)}.
\label{coolA}
\end{equation}
\noindent In this expression, $k_{B}$ is the Boltzmann constant, $T_{LS}$ is the post-shock temperature at the leading shock (calculated by means of the Rankine-Hugoniot relations) and $\Lambda$ is the cooling function for a gas in collisional ionization equilibrium (CIE). In our simulations,  $t_{c}$ is calculated at each time-step and the minimum $t_{min}$ is determined:
\begin{equation}
t_{min}=min \left(t_{c}\left(t \right), t_{c}\left(t+\Delta t \right), \ldots \right).
\label{shellftime}
\end{equation}
The transition time is $t_{sf}=t_{min}$. \\
\indent The evolution of the reverse shock position $R_{RS}$ is calculated as:
\begin{equation}\label{eq:4}
\frac{d R_{RS}}{dt}=\frac{R_{RS}}{t}-\tilde{V}_{RS},
\end{equation}
where $\tilde{V}_{RS}$ is the reverse shock velocity in the frame of the unshocked ejecta. From the Rankine-Hugoniot relations (e.g. \citealt{1995PhRMckee}):
\begin{equation}\label{eq:5}
\tilde{V}_{RS}^{2}=\frac{\gamma+1}{2}\frac{P_{RS}\left(R_{RS},t \right)}{\rho_{ej}\left(R_{RS},t \right)},
\end{equation}
where $P_{RS}\left(R_{RS},t \right)$ is the gas pressure just behind the reverse shock in zone B (Fig. \ref{fig1}) and $\rho_{ej}\left(R_{RS},t \right)$ is the density of the unshocked ejecta  in front of the reverse shock (zone C in Fig. \ref{fig1}). \\
Several calculations (e.g. \citealt{gull1973numerical}, \citealt{gull1975x}, \citealt{chevalier1982radio}, \citealt{hamilton1984new}, \citealt{Silich2018}) have shown that the thermal pressure of the shocked ejecta and the shocked ambient gas in zones B and A rapidly becomes almost homogeneous but presents a sharp fall just behind the reverse shock. Therefore, $P_{RS}$ is smaller than the average pressure of the shocked gas in zones A and B: $P_{RS}<P$. Thus, the pressure ratio $\phi=P_{RS}/P<1$ \citep[hereafter \citetalias{1999Mckee}]{1999Mckee}. For example, $\phi=0.3$ for both steep power-law ejecta \citep{chevalier1982self} and uniform ejecta \citep{hamilton1984new}. As it is shown in Appendix \ref{Ap2}, $\phi \left(t_{0} \right)\approx 0.3$ also for the fiducial initial conditions adopted here. Moreover, as the reverse shock approaches the center of the explosion, $\phi$ also reaches values close to 0.3 \citep{Gaffet1}. Numerical simulations also show that $\phi$ slowly changes with time (e.g. \citealt{Fabian}). Therefore, hereafter the thermal pressure $P$ between $R_{LS}$ and $R_{RS}$ is assumed to be uniform but drops rapidly nears the reverse shock such that $\phi=0.3$. 
\subsection{The evolution of the remnant energies}
In order to solve equations (\ref{eq:1}-\ref{eq:4}), one needs to know the thermal pressure $P$, which is calculated here by means of the energy conservation equation. It is assumed that the ejecta density distribution is given by a power-law of index $n<3$ and that the ejected gas freely expands. Hence, the kinetic energy of the free ejecta is readily integrated (see Appendix \ref{Ap1}):
\begin{equation}\label{eq:7}
E_{k,free}=\frac{1}{2}M_{ej}V_{ej}^{2} \left(\frac{3-n}{5-n} \right)\left(\frac{R_{RS}}{t V_{ej}} \right)^{5-n},
\end{equation} 
where $M_{ej}$ and $V_{ej}$ are the ejecta mass and the maximum expansion velocity, respectively. The kinetic energy of the swept-up ambient gas is:
\begin{equation}\label{eq:8}
E_{k,ism}=\frac{1}{2}M_{s1}U_{s1}^{2}.
\end{equation}
It is assumed that the shocked ejecta moves with the same velocity as the swept-up ambient gas, therefore:
\begin{equation}\label{eq:9}
E_{k,ej}=\frac{1}{2}M_{s2}U_{s1}^{2},
\end{equation}
where $M_{s2}$ is the mass of the thermalized ejecta (see Appendix \ref{Ap1}):
\begin{equation}\label{eq:10}
M_{s2}=M_{ej}\left[1-\left(\frac{R_{RS}}{V_{ej}t} \right)^{3-n} \right].
\end{equation}
The energy conservation equation reads as:
\begin{equation}\label{eq:11a}
E_{0}=E_{th}+E_{k,free}+E_{k,ej}+E_{k,ism}+E_{rad1}+E_{rad2},
\end{equation}
where $E_{rad1}$ and $E_{rad2}$ are energies lost by radiation at the outer and inner shells, respectively. Equation (\ref{eq:11a}) can be written as a differential equation for the thermal energy:
\begin{equation}\label{eq:11}
\frac{d E_{th}}{dt}=-\frac{d E_{k,free}}{dt}-\frac{d E_{k,ej}}{dt}-\frac{d E_{k,ism}}{dt}-Q_{1}-Q_{2},
\end{equation}
where $Q_{1}$ and $Q_{2}$ are the cooling rates in the shocked ambient gas and the shocked ejecta, respectively: 
\begin{equation}\label{eq:12}
Q_{1} = \left\lbrace
\begin{array}{ll}
  n_{s1}^{2}\Lambda \left(T_{LS} \right)\Omega_{s1}, &\text{if} \hspace{0.3cm}  t \leq t_{sf}, \\
  4 \pi R_{LS}^{2}U_{LS}P-\frac{dE_{k,sim}}{dt}, & \text{if} \hspace{0.3cm} t >t_{sf},
\end{array}
\right.
\end{equation}
\begin{equation}\label{eq:15}
Q_{2}= n_{s2}^{2}\Lambda \left(T_{RS} \right) \Omega_{s2}.
\end{equation}
The terms $n_{s1}$, $n_{s2}$ and $\Omega_{s1}$, $\Omega_{s2}$ in equations (\ref{eq:12}-\ref{eq:15}) are the densities and the volumes occupied by the shocked ambient gas and the shocked ejecta:
\begin{equation}\label{eq:13}
n_{s1} = \frac{\rho_{s1}}{\mu m_{H}}=\frac{P}{k_{B}T_{LS}m_{H}}, 
\end{equation}
\begin{equation}\label{eq:14}
\Omega_{s1} = \frac{M_{s1}}{\rho_{s1}}.
\end{equation}
\begin{equation}\label{eq:16}
n_{s2}=\frac{P_{RS}}{k_{B}T_{RS}m_{H}}, 
\end{equation}

\begin{equation}\label{eq:17}
\Omega_{s2}=\frac{M_{s2}}{\rho_{s2}}, \hspace{0.5cm} \rho_{s2}=\mu m_{H} n_{s2}.
\end{equation}
\noindent In these equations, $T_{RS}$ is the post-shock temperature at the reverse shock.\\
\indent The time derivatives of the kinetic energies are calculated with equations (\ref{eq:7}-\ref{eq:10}). Indeed, from equations (\ref{eq:8}-\ref{eq:9}), and by making use of equations (\ref{eq:2}) and (\ref{eq:4}) one can obtain:
\begin{equation}
\frac{d E_{k,ism}}{dt}=4 \pi P R_{LS}^{2}U_{s1}-\frac{1}{2}U_{s1}^{2}\frac{d M_{s1}}{dt},
\end{equation}
\begin{equation}
\frac{d E_{k,ej}}{dt}=\frac{U_{s1}^{2}}{2}\frac{dM_{s2}}{dt}+\frac{M_{s2}}{M_{s1}}U_{s1}\left[\frac{d E_{k,ism}}{dt}-\frac{U_{s1}^{2}}{2}\frac{dM_{s1}}{dt} \right],
\end{equation}
where:
\begin{equation}
\frac{dM_{s2}}{dt}=\left(3-n \right) \frac{M_{ej}}{V_{ej}t} \left( \frac{R_{RS}}{V_{ej}t}\right)^{2-n}\tilde{V}_{RS}.
\end{equation}
From equations (\ref{eq:7}) and (\ref{eq:4}):
\begin{equation}
\frac{d E_{k,free}}{dt}=-\frac{\left(3-n\right)}{2v_{ej}t}M_{ej}V_{ej}^{2}\tilde{V}_{RS}\left(\frac{R_{RS}}{V_{ej}t} \right)^{4-n}.
\end{equation}
The thermal pressure is then calculated as:
\begin{equation}\label{eq:18}
P=\left(\gamma-1 \right) \frac{E_{th}}{\Omega_{LS}-\Omega_{RS}},
\end{equation}
where $\Omega_{LS}$ and $\Omega_{RS}$ are the volumes encompassed by the leading and the reverse shock, respectively.
\subsection{The Initial Conditions}\label{initialc}
The values of $E_{0}$, $M_{ej}$ and $n$ are set at the initial time $t_{0}$. Then, $V_{ej}$ is obtained from equation (\ref{eq:A4}). At $t_{0}$, a small fraction $\beta$ (usually $\beta <5 \%$) of the energy $E_{0}$ is assumed to be already transformed into $E_{k,ism}$, $E_{th}$ and $E_{k,ej}$, i.e:
\begin{equation}\label{eq:19a}
E_{k,free}\left(R_{RS}\left(t_{0} \right), t_{0} \right)=\left(1-\beta \right)E_{0}.
\end{equation}
\begin{figure}
	\includegraphics[width=\columnwidth]{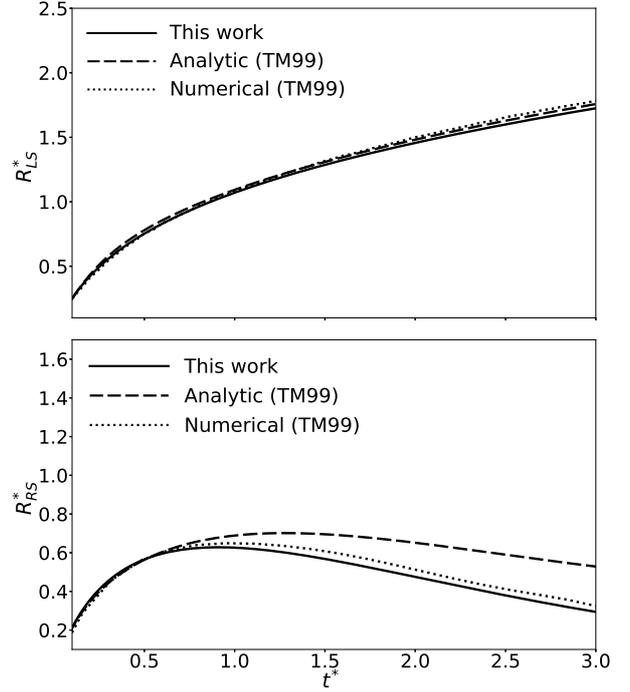}
    \caption{The evolution of the shocks radii for a SNR with $E_{0}=10^{51}$ erg, $n_{0}=1$ cm$^{-3}$, $M_{ej}=3 M_{\odot}$ and index $n=2$. The top panel presents the case of the leading shock $R_{LS}$ and the bottom panel the reverse shock $R_{RS}$. The solid lines show our results and the dashed and dotted lines are the analytic and numerical radii obtained by \citetalias{1999Mckee}, respectively. The starred variables at the axis are dimensionless variables as defined in \citetalias{1999Mckee}. }
    \label{fig2}
\end{figure}
Here, we show that the parameters $E_{0}$, $M_{ej}$, $n$, and $\beta$ define the initial conditions for the further remnant evolution.\\
\indent At the initial position of the reverse shock $R_{RS}\left(t_{0} \right)$, the free ejecta velocity is:
\begin{equation}
V_{0}=\frac{R_{RS}\left( t_{0}\right)}{t_{0}}=V_{RS}\left(t_{0} \right).
\end{equation}
In order to determine $V_{0}$, one can make use of the equations (\ref{eq:7}) and (\ref{eq:19a}):
\begin{equation}\label{eq:19}
\left(1-\beta \right)E_{0}=\frac{1}{2}M_{ej}V_{ej}^{2} \left(\frac{3-n}{5-n} \right)\left(\frac{V_{0}}{V_{ej}} \right)^{5-n}.
\end{equation}
This equation together with equation (\ref{eq:A4}) from Appendix \ref{Ap1}  yield: 
\begin{equation}\label{eq:20}
V_{0}=V_{ej}\left(1-\beta \right)^{1/\left(5-n \right)},
\end{equation}
where $V_{ej}$ is the maximum expansion velocity of the ejecta. Following  \cite{chevalier1982self}, \cite{hamilton1984new}, \cite{Hwang2012a}, we define the leading factor $l_{ED}$ at $t_{0}$ as:
\begin{equation}\label{eq:21}
l_{ED}= \frac{R_{LS}\left( t_{0} \right)}{R_{RS}\left( t_{0} \right)}.
\end{equation}
The relation between the initial leading and reverse shock radii then is:
\begin{equation}\label{eq:22}
R_{LS}\left( t_{0} \right)=l_{ED}R_{RS}\left( t_{0} \right).
\end{equation} 
\begin{figure*}
	\includegraphics[width=1.8\columnwidth]{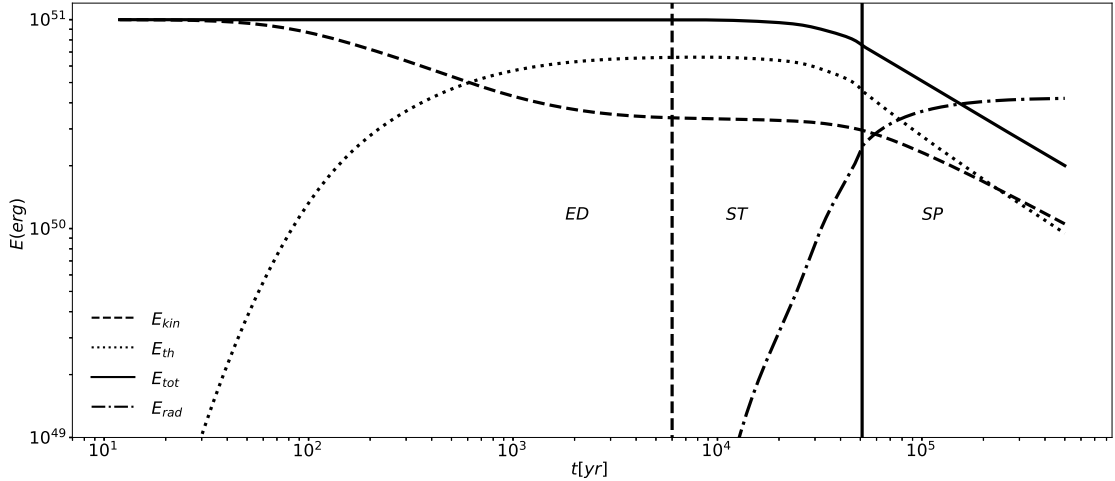}
    \caption{Evolution of the total, thermal, kinetic and radiated energies (solid, dotted, dashed, and dash-dotted lines, respectively) for a SNR evolving in the ambient medium with density $n_{0}=1$ cm$^{-3}$. The three evolutionary stages are separated by the dashed and solid vertical lines. The dashed vertical line marks the moment when the reverse shock reaches the center of the explosion and the solid vertical line the thin-shell formation time (see section \ref{massConserva}). }
    \label{fig3}
\end{figure*}
The position of the reverse shock is assumed to be coincident with the contact discontinuity (\citetalias{1999Mckee}, see left panel of Fig. \ref{fig1}). One can obtain then the value of $l_{ED}=1.1$ from the mass conservation equation. From equation (\ref{eq:22}):
\begin{equation}\label{eq:23}
V_{LS}\left( t_{0} \right)=l_{ED}V_{0},
\end{equation}
where $V_{LS}\left( t_{0} \right)$ is the velocity of the leading shock at $t_{0}$. The initial velocity of the shocked ambient gas then is:
\begin{equation}\label{eq:24}
U_{s1}\left(t_{0} \right)=\frac{2}{\gamma+1}l_{ED}V_{0}.
\end{equation}
Finally, the shock positions are determined from equation (\ref{eq:22}) and the energy conservation equation. Indeed, as $\beta$ is the fraction of the ejecta kinetic energy converted into other energies, then:
\begin{equation}\label{eq:25}
\beta E_{0}= E^{0}_{th}+E^{0}_{k,ej}+E^{0}_{k,ism},
\end{equation}
where the terms on the right-hand side of this expression are the thermal and kinetic energies of the shocked ejecta and the shocked ambient gas at $t_{0}$:
\begin{equation}\label{eq:26}
E^{0}_{k,ism}= \frac{1}{2}\rho_{0}\frac{4 \pi}{3}R_{LS}^{3}\left( t_{0} \right)\left( \frac{2}{\gamma+1}l_{ED}V_{0}\right)^{2},
\end{equation}
\begin{equation}\label{eq:27}
E^{0}_{k,ej}=\frac{1}{2}M_{ej}\left[1-\left(\frac{V_{0}}{V_{ej}}\right)^{3-n} \right]\left( \frac{2}{\gamma+1}l_{ED}V_{0}\right)^{2},
\end{equation}

\begin{equation}\label{eq:28}
E^{0}_{th}=\frac{4}{\gamma-1}\frac{k_{B}\rho_{0}}{\mu} T_{LS} \frac{4\pi}{3}\left(1-\frac{1}{l_{ED}^{3}} \right)R_{LS}^{3}\left( t_{0} \right).
\end{equation}
Equation (\ref{eq:28}) was derived under the assumption that the post-shock density is $4 n_{0}$. The initial position for the leading shock radius $R_{LS}\left( t_{0} \right)$ is calculated from equation (\ref{eq:25}) as this is the only unknown parameter in equations (\ref{eq:26}-\ref{eq:28}). Finally, the initial time $t_{0}$ is:
\begin{equation}
t_{0}=\frac{R_{RS}\left(t_{0} \right)}{V_{0}},
\end{equation}
where $R_{RS}\left( t_{0} \right)$ is determined by means of equation (\ref{eq:22}). 
\begin{figure*}
	\includegraphics[width=2\columnwidth]{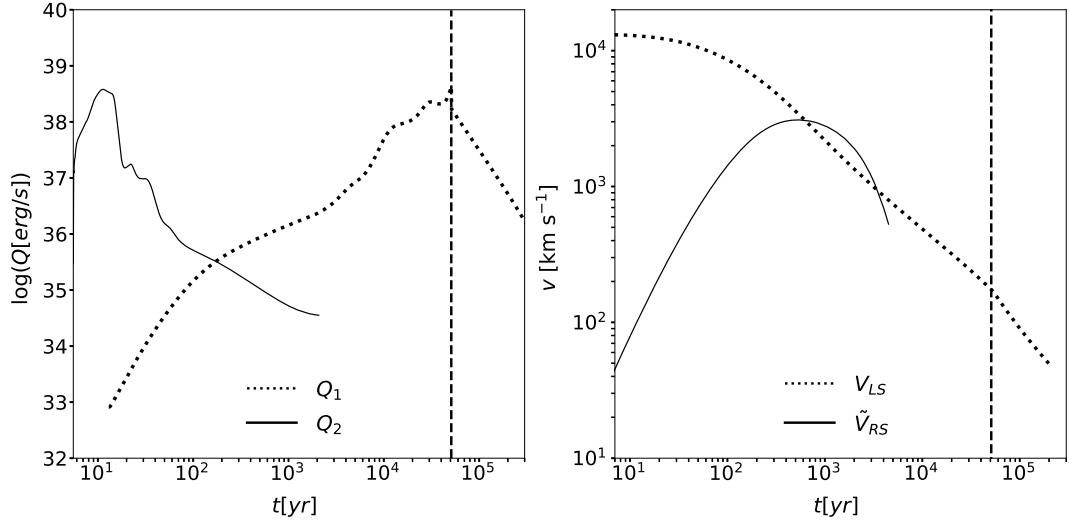}
    \caption{The rate of energy loss (left panel) from the outer (dotted line) and inner shells (solid line) for the case of $n_{0}=1$ cm$^{-3}$. Right panel: the leading (dotted line) and reverse (solid line) shocks velocities as function of time. Note that $\tilde{V}_{RS}$ is the reverse shock velocity in the frame of the unshocked ejecta (see equation \ref{eq:4}) and $V_{LS}$ is the leading shock velocity in the rest frame. The vertical lines indicate the swept-up gas cooling time.}.
    \label{fig4}
\end{figure*}
\section{Comparison to numerical and analytic models}\label{sec3}
In order to integrate equations (\ref{eq:1}-\ref{eq:4}) and (\ref{eq:11}), we have used the Dormand-Prince method for the eight order Runge-Kutta-Fehlberg integrator, coupled with a PI step-size control \citep{press2007numerical}. The cooling was included trough a table lookup/interpolation method. In all calculations, we have used the cooling function from \cite{Raymond1976} for a solar metallicity, unless otherwise stated.
\subsection{SNR evolution in a low density ambient medium}
Fig. \ref{fig2} presents the evolution of the leading (upper panel) and reverse (bottom panel) shock radii  predicted by our model (solid lines) and compare them to the analytic and numerical solutions obtained by \citetalias{1999Mckee} (dashed and dotted lines) in the case when $E_{0}=10^{51}$ erg, $M_{ej}=3 M\odot$, $n_{0}= 1$ cm$^{-3}$ and $n=2$. Following \citetalias{1999Mckee}, the time and shock radii in Fig. \ref{fig2} are presented in the dimensionless form $t^{*}=t/t_{ch}$, $R_{RS}^{*}=R_{RS}/R_{ch}$, $R_{LS}^{*}=R_{LS}/R_{ch}$, where:

\begin{equation}\label{char2}
R_{ch}=M_{ej}^{1/3}\rho_{0}^{-1/3},
\end{equation}
\begin{equation}\label{char3}
t_{ch}=E^{-1/2}M_{ej}^{5/6}\rho_{0}^{-1/3}.
\end{equation}
\indent \citetalias{1999Mckee} studied analytically and numerically the ED and the ST stages for low density media. Cooling was assumed to be negligible and therefore their model is adiabatic.  \\
\indent Our results are in excellent agreement with the numerical results obtained by \citetalias{1999Mckee} both for the leading (see the upper panel in Fig. \ref{fig2}) and the reverse (see lower panel in Fig. \ref{fig2}) shocks. Small differences between ours and \citetalias{1999Mckee} numerical calculations are likely to be produced because in our calculations $\phi$ was assumed to have a constant value while in \citetalias{1999Mckee} numerical calculations $\phi$ slightly changes with time. Note that the reverse shock positions obtained in all cases coincide very well at the ED stage. However, after that the analytic solution (dashed line) is not able to reproduce the correct position of the reverse shock which, as noticed by \citetalias{1999Mckee}, results from a significant error in the reverse shock velocity around the Sedov-Taylor transition time $t_{ST}$. \\
\indent Fig. \ref{fig3} presents the evolution of the remnant energies. The three evolutionary stages are separated by vertical lines. As one can see, during the ED phase, the kinetic energy of the free ejecta is converted into kinetic and thermal energies of the shocked gas. The dashed vertical line marks the beginning of the ST stage. At this time, the ratio of the ambient swept up mass to the ejecta mass is about 38, which is in good agreement with the results of \cite{gull1973numerical} and \cite{1990MNRASTenorio}. At this moment, the energy lost by radiation, $E_{rad}=\int \left(Q_{1}+Q_{2} \right)dt$ (dash-dotted line), is negligible, thus allowing the total kinetic and thermal energies to approximately reach constant values ($E_{k} = E_{k,free}+E_{k,ej}+E_{k,ism} \approx 0.33 E_{0}$ and $E_{th}\approx 0.66 E_{0}$). The leading shock radius then evolves as $R_{LS} \propto t^{0.39}$, which is close to the analytic solution (\citealt{sedov1993similarity}). \\
\indent Note that $E_{rad}$ steadily grows several orders of magnitude to finally make an impact on the evolution terminating the ST stage.  The vertical solid line in Fig. \ref{fig3} marks the transition to the SP phase at $t_{sf}$ (see \ref{massConserva}). For the case considered here, $t_{sf} \approx 5 \times 10^{4}$ yr, which is in agreement with recent results (e.g. \citealt{LiOstriker, KimOstriker, Haid2016}). \\
\indent The left panel in Fig. \ref{fig4} presents the energy loss rate behind the leading ($Q_{1}$) and the reverse ($Q_{2}$) shocks. The right panel shows the reverse shock velocity $\tilde{V}_{RS}$ in the frame of the unshocked ejecta (see equation \ref{eq:4}) and the velocity of the leading shock $V_{LS}$ in the rest frame, as a function of time. The vertical dashed lines on both plots indicate the thin-shell formation time.  At the beginning of the evolution, the reverse shock is radiative as the ejecta density is large and the reverse shock velocity $\tilde{V}_{RS}$ is small (solid line on the right panel). However, this period is short because the shock velocity $\tilde{V}_{RS}$ grows and the ejecta density drops. Hence, the radiative losses $Q_{2}$ become negligible for most of the evolution. On the other hand, the leading shock velocity is high at early times (dotted line on the right panel). This implies that $Q_{1}$ is initially small but continuously increases as the shock slows down. When the post-shock temperature drops to values close to the maximum of the cooling function, $Q_{1}$ reaches the maximum value. Note how close this luminosity peak is to $t_{sf}$. This is in agreement with previous results by \cite{Thornton1998}, who used the luminosity peak as the definition of $t_{sf}$.\\
\begin{figure}
	\includegraphics[width=\columnwidth]{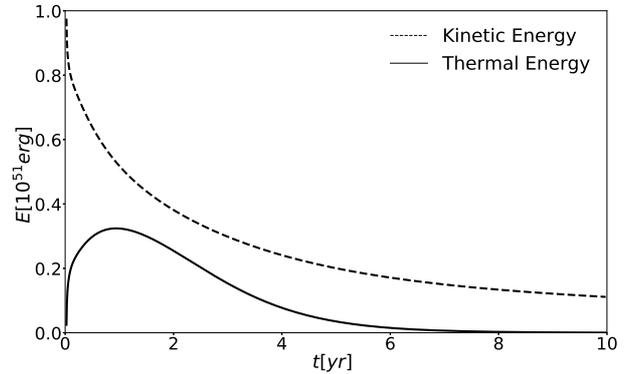}
    \caption{The kinetic (dashed line) and thermal (solid line) energies of a SNR evolving in an ambient medium with density $n_{0}=10^{7}$ cm$^{-3}$. }
    \label{fig5}
\end{figure}
\begin{figure*}
	\includegraphics[width=2\columnwidth]{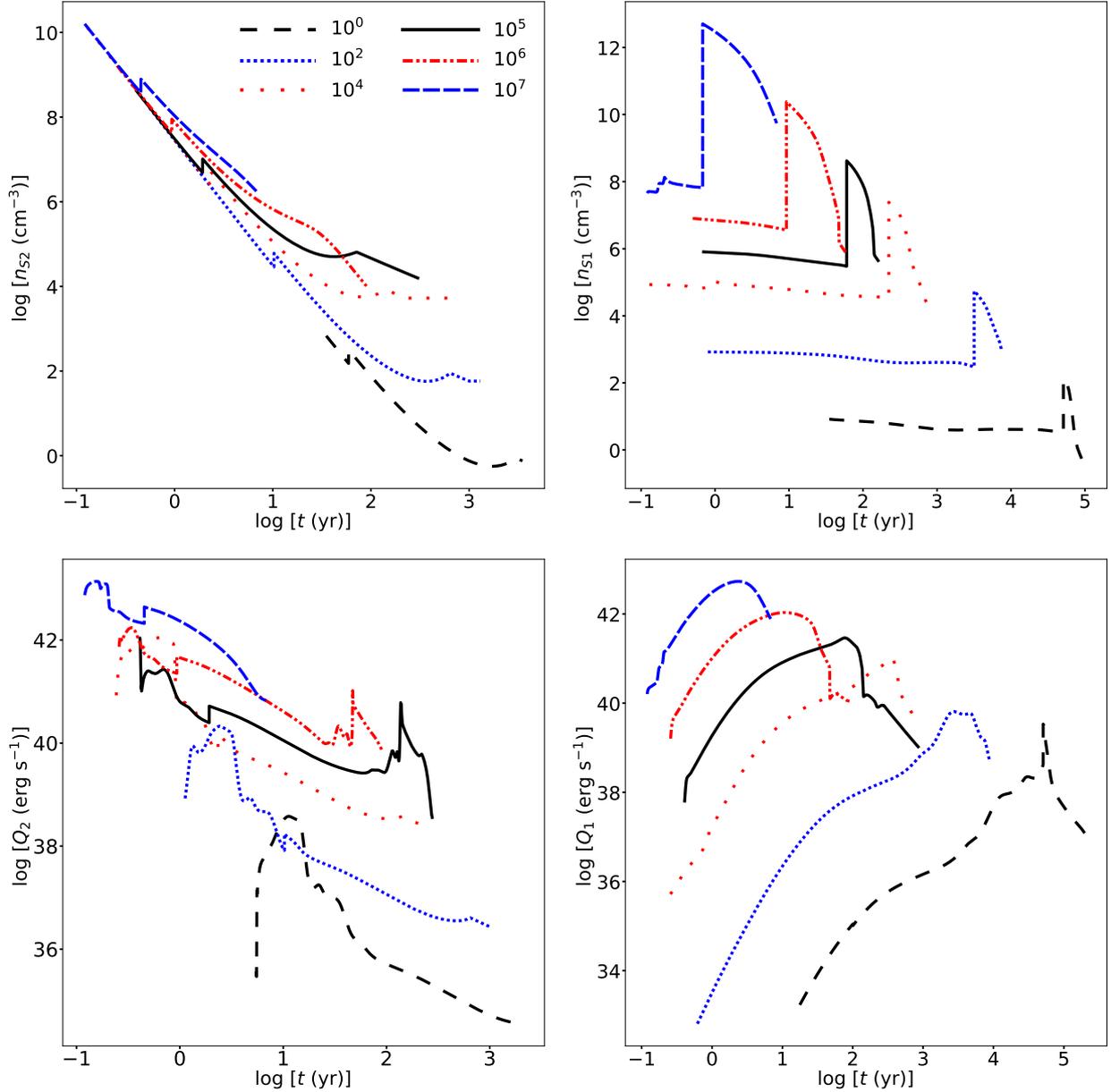}
    \caption{The shocked ambient gas and the shocked ejecta densities and the cooling rates as a function of time. The left and right upper panels present the shocked ejecta $n_{s2}$ and shocked ambient gas $n_{s1}$ densities for different ambient gas densities (shown in the legend in units of cm$^{-3}$). The left and right lower panels present the corresponding rates of energy losses $Q_{2}$ and $Q_{1}$.}.
    \label{fig6b}
\end{figure*}
\subsection{SNR evolution in a high density medium}
A test case for the SNR evolution in a high density medium was presented in \cite{terlevich1992starburst}, who discussed the SNR evolution in a $n_{0}=10^{7}$ cm$^{-3}$ ambient medium. To compare our model with these results, the velocity and density distributions of the ejected gas and the initial conditions were modified (see Appendix \ref{Ap3}) to account for the initial values used by \cite{terlevich1992starburst}. Fig. \ref{fig5} presents the evolution of the remnant energies for this case. The strong radiative cooling begins to be a dominant factor around $\approx 1 $ yr after the explosion, which is the time when the leading shock becomes radiative. This leads to the rapid remnant evolution as the thermal energy dramatically decreases in a short time interval. The fact that the remnant energies are decaying for most of the time covered by our calculations implies that the energy of the newly shocked gas is radiated very efficiently. The total kinetic and thermal energies never reach the values of $E_{k} \approx 0.33 E_{0}$ and $E_{th} \approx 0.66 E_{0}$ and therefore the Sedov-Taylor stage is inhibited and radiative cooling sets in before the thermalization is completed, in agreement with the numerical results presented by \cite{terlevich1992starburst}. 

\begin{figure}
	\includegraphics[width=\columnwidth]{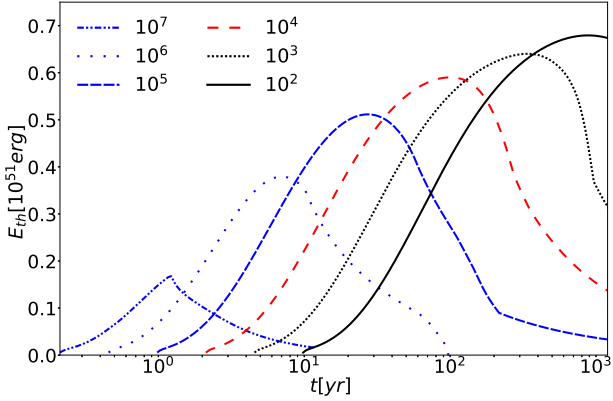}
    \caption{The thermal energy of SNRs evolving into an ISM with different densities (listed in the legend in units of cm$^{-3}$). }
    \label{fig6}
\end{figure}



\section{SNR Evolution in different ambient media}\label{sec4}
\subsection{The impact of the ambient gas density}\label{sec4-1}
Our numerical model allows one to study the evolution of SNRs in a wide range of ambient gas densities. This is problematic if one uses full hydrodynamical codes, because such calculations require high spatial and temporal resolutions and therefore are time-consuming (e.g. \citealt{leveque2006computational}). This section presents the results of simulations which were provided for ambient gas densities: $n_{0}$ $[$cm$^{-3}]=10^{2},$ $10^{3},$ $5 \times 10^{3},$ $10^{4},$ $10^{5},$ $5 \times 10^{5},$ $10^{6},$ and $10^{7}$. All calculations assume that the ejecta mass is $M_{ej}=3$ $M_{\odot}$, the total energy is $E_{0}=10^{51}$ erg and the ejecta density distribution is a power-law with index $n=2$. \\
\begin{figure}
	\includegraphics[width=\columnwidth]{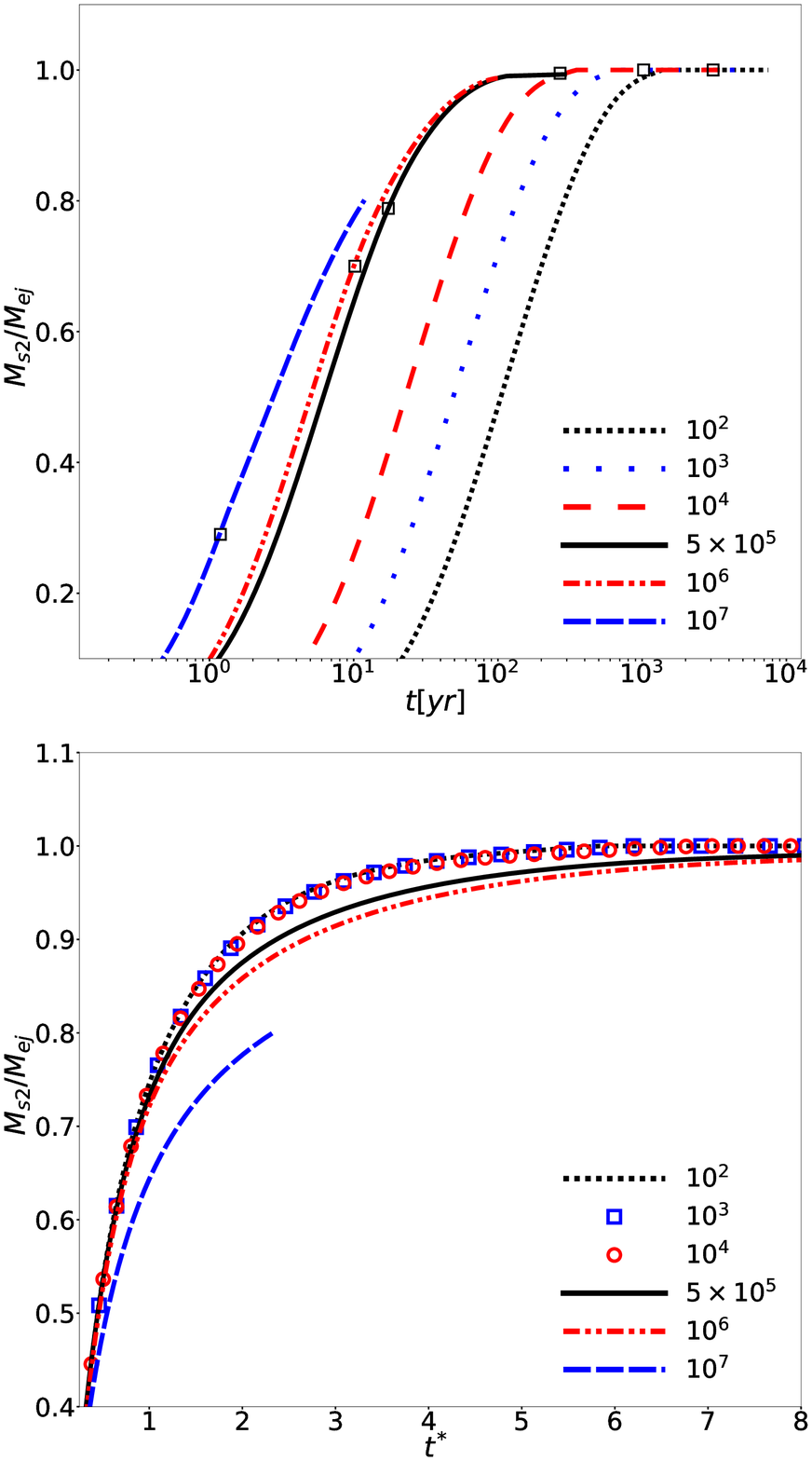}
    \caption{Top panel: Fraction of the thermalized ejecta $M_{s2}/M_{ej}$ as a function of time  for different values of the ambient gas density (shown in the legend). The open squares indicate the time when the leading shock becomes radiative (i.e., when $t=t_{sf}$) for each case. For low density models this occur after full thermalization of the ejecta ($M_{s2}/M_{ej}=1$), in contrast with the large density models when $M_{s2}/M_{ej}<1$  at $t_{sf}$. Bottom panel: Same as the top panel but the time coordinate is now expressed in dimensionless units (see Appendix \ref{Ap1}). Unified solutions must lie over the same curve in this plot. This is the case for densities $n_{0}<5 \times 10^{5}$ cm$^{-3}$ but higher densities cases depart from this curve. }
    \label{fig6a}
\end{figure}
\begin{figure*}
	\includegraphics[width=2\columnwidth]{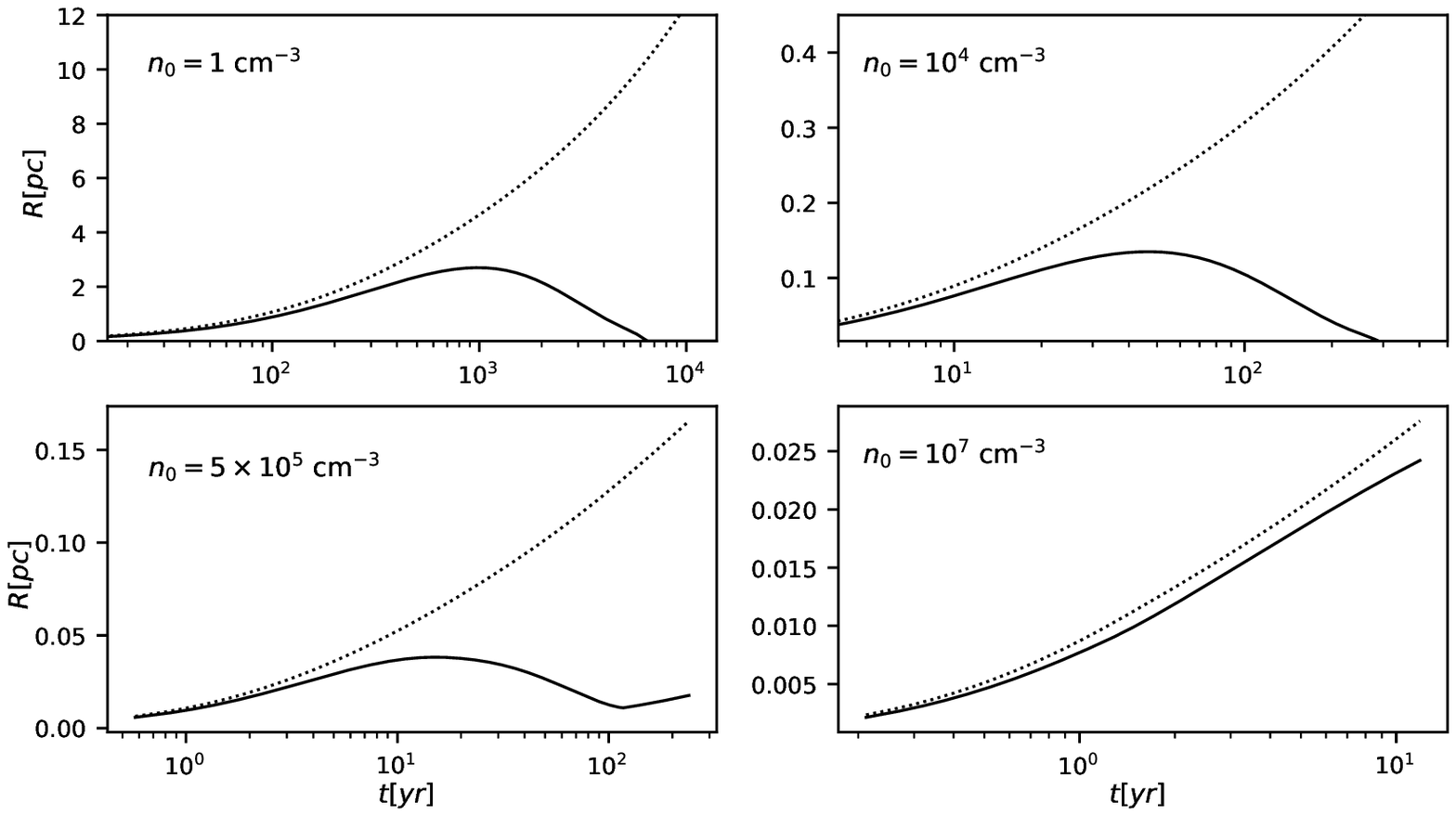}
    \caption{Evolution of the reverse (solid line) and leading (dotted line) shock radii for the densities $n_{0}=1, 10^{4}, 5 \times 10^{5}$ and $10^{7}$ cm$^{-3}$, respectively.}
    \label{fig8}
\end{figure*}
\indent Fig. \ref{fig6b} presents the shocked ejecta and the shocked ambient gas densities ($n_{s2}$ and $n_{s1}$)  and cooling rates ($Q_{2}$, $Q_{1}$) for different ambient gas densities as functions of time. One can note that the initial value of $n_{s2}$ is large in all cases. However, it rapidly drops with time, slightly increasing at the end of the ED stage as the ejecta density is large near the center of the explosion. The shocked ejecta density $n_{s2}$ reaches smaller values in calculations with smaller ambient gas densities as in these cases the ejecta gas passes through the reverse shock at larger distances from the center of the explosion. The density of the shocked ambient gas $n_{s1}$ (right upper panel) is initially close to the adiabatic strong shock limit $4 n_{0}$. However, it increases orders of magnitude upon strong radiative cooling at the transition time $t_{sf}$. At final stages of the SNRs evolution, $n_{s1}$ falls because radiative cooling becomes inefficient as the leading shock decelerates. \\
\indent The left bottom panel in Fig. \ref{fig6b} presents the evolution of $Q_{2}$. In large ambient gas densities strong radiative cooling sets in earlier and $Q_{2}$ reaches larger maximum values than in low ambient gas densities. This occurs because in these models the density of the shocked ejecta is larger. $Q_{1}$ presents a similar trend: it reaches larger maximum values at earlier evolutionary times in models with larger ambient gas densities (see the right bottom panel in Fig. \ref{fig6b}). This is due to larger shocked ambient gas densities $n_{s1}$ (see right upper panel). Note that in the high-density models, the cooling rates $Q_{1}$ and $Q_{2}$ reach their maximum values at similar times whereas in the low density models this does not occur.  \\
\indent Fig. \ref{fig6} shows the evolution of the thermal energy for various values of $n_{0}$. As discussed before, in the large ambient density cases radiative cooling sets in at early stages of the SNR evolution and therefore $E_{th}$ never reaches values predicted by the ST solution. \\
\indent In low density media, thermalization is well separated from the radiative stage by the Sedov-Taylor regime. However, one can notice from our calculations that in high density media the leading shock becomes radiative before the thermalization is completed. This implies that in these cases both processes (the ejecta thermalization and the shocked gas cooling) proceed simultaneously. Indeed, the top panel of Fig. \ref{fig6a} presents the fraction of thermalized ejecta $M_{s2}/M_{ej}$ as a function of time. The open squares mark this fraction at the time when the leading shock becomes radiative (i.e, $t=t_{sf}$). For densities $n_{0}<5 \times 10^{5}$ cm$^{-3}$, practically all the ejected gas is thermalized by this time, while in the larger ambient gas densities this fraction becomes progressively smaller. \\
\indent As has been shown by \cite{1999Mckee}, \cite{2008A&A...478...17F}, \cite{2009MNRAS.397.2106T}, in low density media the leading and reverse shock radii and velocities scale with the input parameters $E_{0}$, $n_{0}$ and $M_{ej}$. This implies that the radii and velocities are determined by a unified dimensionless solution. Equation (\ref{Ap:terma}) shows that for a given power-law index $n$, the thermalized ejecta mass $M_{s2}/M_{ej}$ is also a function of the dimensionless variables $t^{*}$ and $R^{*}_{RS}$ (\citetalias{1999Mckee}). The bottom panel on Fig. \ref{fig6a} presents the ratio $M_{s2}/M_{ej}$ as a function of $t^{*}$. As both axes are now dimensionless, the ratio $M_{s2}/M_{ej}$ should not change with the ambient gas density $n_{0}$ if a unified solution exists. Indeed, this is the case for low densities ($n_{0}<5 \times 10^{5}$ cm$^{-3}$) but is not true in denser ambient media. Therefore, the early radiative cooling in high density media breaks the scaling properties of the SNRs evolution. \\
\indent The radiative cooling impact the reverse shock dynamics. For low density cases, the reverse shock promptly reaches the center of the explosion (see upper panels of Fig. \ref{fig8}) as the thermal pressure of the shocked gas is always larger than the ejecta ram pressure, thus allowing the SNRs to thermalize all of its ejected mass before the onset of the SP stage. However, in the high density cases, the thermal pressure drops drastically and becomes smaller than the ejecta ram pressure. Our calculations show that for densities larger than the critical density $n_{0,cri}=5 \times 10^{5}$ cm$^{-3}$, the reverse shock never reaches the center of the explosion and instead moves outwards (see bottom panels of the Fig. \ref{fig8}). In these cases the leading shock decelerates rapidly and this leads to the merging of the shells of shocked ejecta and ambient gas, shortly after the explosion (see the bottom right panel of \ref{fig8}, where the dotted line indicates the leading shock position).  \\
\begin{figure*}
	\includegraphics[width=1.75\columnwidth]{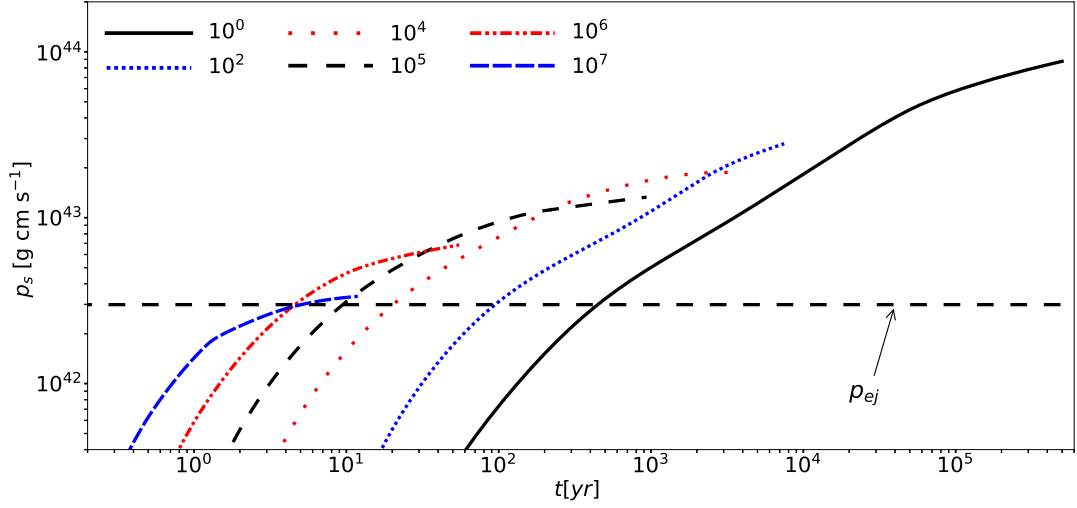}
    \caption{The total momentum deposited by the shocked gas for different values of the ambient gas density. The horizontal line shows the initial momentum of the ejecta $p_{ej}$.}
    \label{fig7}
\end{figure*}
\indent Fig. \ref{fig7} presents the  momentum carried by the shocked gas (i.e. the shocked ejecta and the shocked ambient gas) for several values of the ambient gas density. The horizontal line is the initial momentum of the ejecta $p_{ej}\approx 3.0\times 10^{42}$ g cm s$^{-1}$ (see Appendix \ref{Ap1}). Note that for the same values of the explosion energy and the ambient gas density $n_{0}=1$ cm$^{-3}$, \cite{LiOstriker} obtained numerically a final momentum of $p_{s} \approx 5.0 \times 10^{43}$ g cm s$^{-1}$ while our calculations lead to $p_{s} \approx 8.8 \times 10^{43}$ g cm s$^{-1}$. Note also that in all cases the remnants momentum asymptotically approaches a constant final value $p_{f}$ at the end of the calculations. This is because the thermal pressure of the shocked gas becomes negligible due to cooling and therefore at late stages the SNR evolves in a momentum conservation regime (see equation \ref{eq:2}). \\
\indent Fig. \ref{fig7} also shows that the work done by the shocked gas cannot boost the momentum injected by the explosion too much ($1$ $\leq p_{s}/p_{ej} \leq$ $29$) and that the boosting factor becomes negligible for SNRs evolving in high density media as in these cases the shocked gas cools down rapidly and the SNRs evolve practically in the momentum-dominated regime. This agrees with recent results by \cite{2013ApJ...770...25A}, \cite{2015MNRAS.450..504M} and \cite{2015MNRAS.451.2757W}, who studied the SNRs evolution for the density range $1$ cm$^{-3}$ $\leq n_{0} \leq$ $100$ cm$^{-3}$. 
\subsection{The impact of the gas metallicity}
\begin{figure*}
	\includegraphics[width=2\columnwidth]{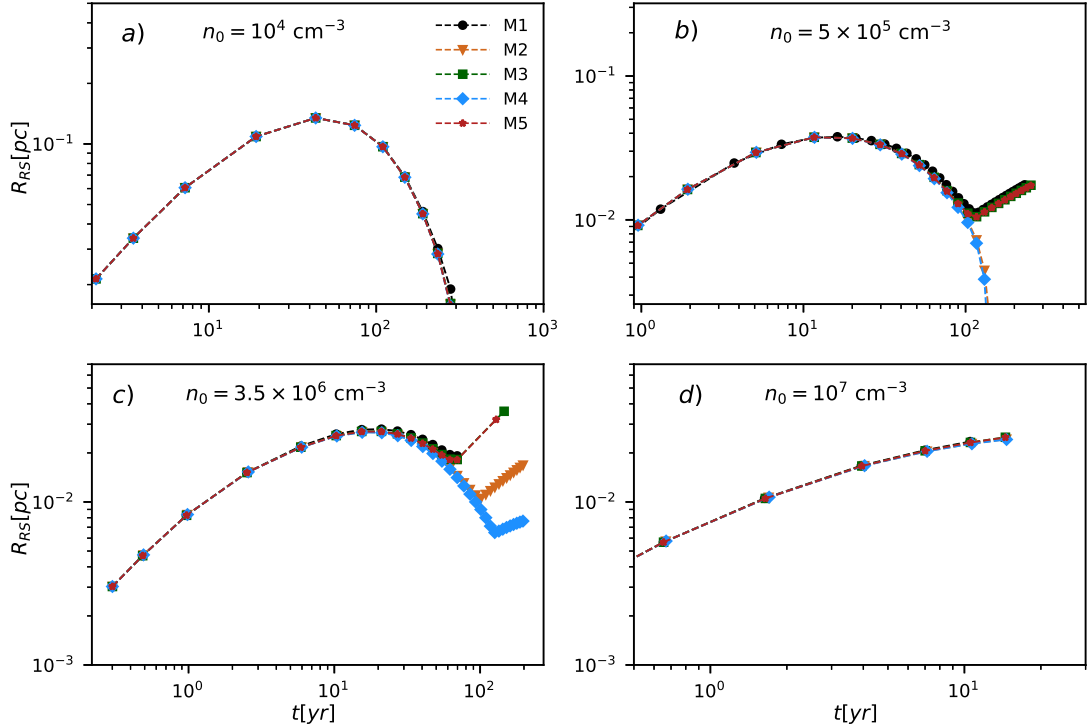}
    \caption{The evolution of the reverse shock position for models with different chemical compositions. Panels a, b,c, and d present the reverse shock position as a function of time for ambient gas densities $10^{4}$ cm$^{-3}$, $5 \times 10^{5}$ cm$^{-3}$, $3.5 \times 10^{6}$ cm$^{-3}$ and $10^{7}$ cm$^{-3}$, respectively.}
    \label{fig7a}
\end{figure*}
\begin{figure*}
	\includegraphics[width=2\columnwidth]{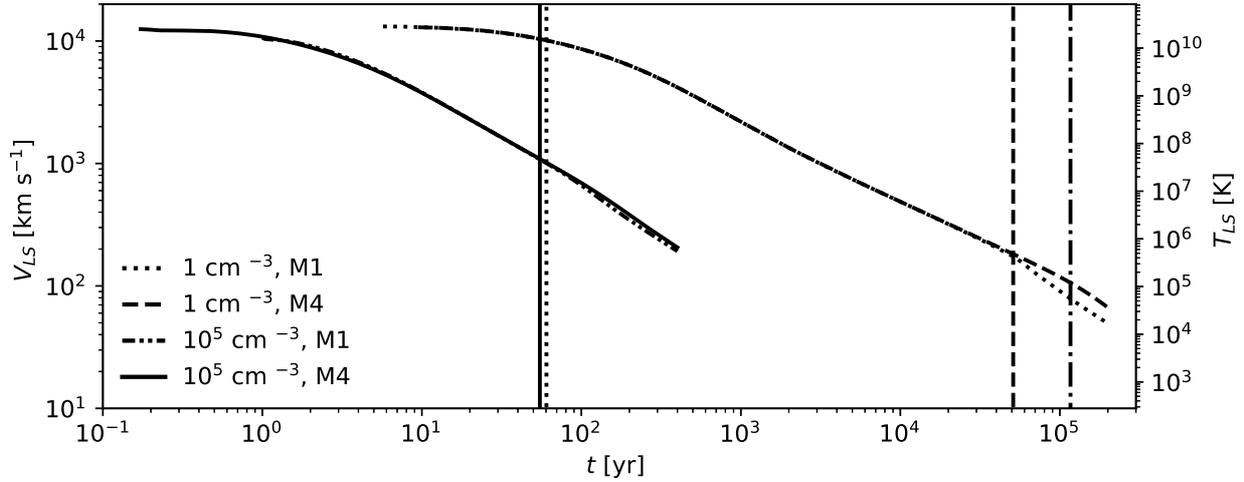}
    \caption{The time evolution of the leading shock velocity $V_{LS}$ for the cases presented in the legend. The right $y$-axis shows the post-shock temperature $T_{LS}$. The solid, dotted, dashed and dash-dotted vertical lines present the thin-shell formation time $t_{sf}$ for the cases ($10^{5}$ cm$^{-3}$, M1), ($10^{5}$ cm$^{-3}$, M4), ($1$ cm$^{-3}$, M1) and ($1$ cm$^{-3}$, M4), respectively.}
    \label{fig7b}
\end{figure*}

The calculations presented in the previous sections  assumed a solar metellicity for the ambient gas and for the SN ejecta. Here, models with $M_{ej}=3 M_{\odot}$, $E_{0}=10^{51}$ erg , $n=2$ and lower metallicities are discussed (see Table \ref{TableMetal}). Note that model $M1$ is the one discussed in previous sections. \\
\indent Fig. \ref{fig7a} presents the evolution of the reverse shock for all these models in different ambient gas densities. In the case of $n_{0}=10^{4}$ cm$^{-3}$ shown in the upper left panel, and in lower density cases, the thermalization of the ejecta gas occurs before radiative cooling becomes important and therefore the reverse shock position practically does not depend on the gas metallicity. The right upper panel shows that in the case $n_{0}=5 \times 10^{5}$ cm$^{-3}$, the reverse shock reaches the center of the explosion only in the low-metallicity models M2 and M4. However, a slight increase in $n_{0}$ to $3.5 \times 10^{6}$ cm$^{-3}$ leads in all cases, to reverse shocks unable to reach the center (see the bottom left panel in Fig. \ref{fig7a}). \\
\begin{table}
\caption{Set of models with different gas compositions. Left panel: model identifier. Second and third columns: the gas metallicity for the ejecta and the ambient gas, respectively.  }
\begin{center}
\begin{tabular}{c c c }
\toprule
\toprule
Model Reference &\multicolumn{2}{c}{Gas Metallicity [$Z_{\odot}$]}\\
\toprule
& Ejecta& ISM\\
\toprule
M1& 1 & 1\\
M2& $10^{-1}$ & $10^{-1}$\\
M3& 1 & $10^{-1}$\\
M4& $10^{-2}$ & $10^{-2}$\\
M5& 1 & $10^{-2}$\\
\toprule
\end{tabular}
\end{center}
\label{TableMetal}
\end{table}
\indent The bottom panels in Fig. \ref{fig7a} show that the variation on the gas composition may alter the shock dynamics as the cooling rates change accordingly, but the impact of the ambient gas metallicity is less important than the ambient gas density (see models M1, M3 and M5 in Fig. \ref{fig7a}). This is due to the fact that free-free cooling dominates the energy losses in SNRs that evolve in large ambient gas densities. To clarify this point, Fig. \ref{fig7b} shows the evolution of the leading shock velocity for a low ($n_{0}=1$ cm$^{-3}$) and a high ($n_{0}=10^{5}$ cm$^{-3}$) density media for models M1 and M4, which are the cases with the highest and lowest metallicities. Fig. \ref{fig7b} also presents the post-shock temperature $T_{LS}$ at the right $y$-axis. The vertical lines indicate $t_{sf}$ for each case. In the low density $n_{0}=1$ cm$^{-3}$, the transition to the SP stage occurs when the post-shock temperatures are $4.3 \times 10^{5}$ K and $ 1.2 \times 10^{5}$ K for M1 and M4, respectively. These temperatures are well within the line-cooling regime. Note that $t_{sf}$ is considerably larger in M4 than in M1 because the shocked ambient gas requires a larger time to cool down in the lower metallicity case. For $n_{0}=10^{5}$ cm$^{-3}$, $t_{sf}$ is similar in both, M1 and M4 models. In this case, the temperatures at $t_{sf}$ are still high ($T_{LS}\approx 4.6 \times 10^{7}$ K and $T_{LS}\approx 4.0 \times 10^{7}$ K for M1 and M4, respectively). At these temperatures, the gas cools mostly due to free-free emission, which is less sensitive to the gas metallicity. Therefore SNRs in low density media cool down as a consequence of the post-shock temperatures reaching values close to the maximum of the cooling function while in high density media, SNRs cool mostly due to the $n_{shock}^{2}$ dependence.

\section{Summary and Discussion}\label{sec5}
A numerical scheme based on the Thin-Shell approximation, which allows one to study the full evolution of SNRs, from the early Ejecta-dominated to the Snowplough stages, has been developed and confronted with a number of previous results. The scheme accounts for the ejecta density and velocity distributions, and for the radiative cooling of the shocked gas. The initial shock radii and velocities are obtained from the explosion parameters: ejecta mass $M_{ej}$, total energy $E_{0}$, power-law index $n$ of the ejecta density distribution and the ambient gas density $n_{0}$, once the fraction of the ejecta kinetic energy $\beta$ that has been thermalized at the initial time $t_{0}$, is selected.  \\
\indent Our model was compared with several previous simulations and reproduces well the evolution of the expansion radii, the remnant energetics, and the momentum deposited by SNRs both in low and high density cases. \\
\indent Our calculations show that radiative cooling speeds up drastically the evolution of SNRs in high density media. In these cases, the thermal energy of the remnants reaches lower maximum values as one considers larger densities. This limits the SNR lifetime and the feedback that SNe provide to the ambient gas.\\ 
\indent It was shown that in high density cases the leading shock becomes radiative long before the thermalization of the ejecta is completed. Therefore in these cases the SNRs never reach the Sedov-Taylor stage, in contrast with the predictions of the standard theory. This implies that the Sedov-Taylor solution cannot be used as the initial condition for numerical simulations of the SNR evolution in high density cases.\\
\indent Strong radiative cooling also impacts the reverse shock dynamics. For densities $n_{0}>10^{5}$ cm$^{-3}$, the thermal pressure falls faster than the ejecta ram pressure and therefore, the reverse shock does not reach the center of the explosion and it is weaker compared to low-density cases. As a consequence, we have shown that scaling relations for the SNRs dynamical evolution are only applicable for $n_{0}<10^{5}$ cm$^{-3}$. \\
\indent The work done by the hot shocked gas increases the momentum deposited by a SN explosion into the ambient medium, although radiative cooling constrains the boosting factor. The lowest boosting factor ($p_{s}/p_{ej} \approx 1$), which corresponds to the highest density medium, implies that in this case the SNR evolves in a momentum conservation regime during most of its evolution, and that the Sedov-Taylor stage is inhibited.\\
\indent The impact of the gas metallicity on the evolution of SNRs was also addressed. Several models with sub-solar compositions for the ejecta and the ambient gas were discussed. In low-density media, the SP stage begins when the post-shock temperatures reach values close to the maximum of the cooling function and therefore the onset of the SP stage depends on the gas metallicity. In high density models, however, the impact of the ambient gas metallicity is small as in these cases most of the energy is radiated away in the free-free emission regime. Hence, regardless of the metallicity, the density $n_{0,cri} \approx 5 \times 10^{5}$ cm$^{-3}$ is still the approximate value that inhibits the Sedov-Taylor phase.\\
\indent The results obtained here can contribute to the understanding of dust formation and evolution at early stages of the Universe, when most of the dust grains are expected to be formed in the supernovae ejecta (e.g. \citealt{2001Todini},  \citealt{2006Marchenko}, \citealt{Bianchi2007}, \citealt{2013ApJ...778..159T}). Indeed, recent calculations suggest that in low ambient gas densities ($n_{0}<100$ cm$^{-3}$), just a small fraction (about $10\%-20 \% $) of dust grains could survive crossing the reverse shock (e.g. \citealt{Bianchi2007}, \citealt{Micelotta2016}) due to thermal sputtering (e.g. \citealt{1979ApJ...231...77D},  \citealt{2017MNRAS.468.1505M}). Our results suggest that a larger fraction of dust grains could survive in SNRs which evolve in high density media as in these cases the reverse shock is weak and the post-shock temperature drops in short timescales, limiting the window of opportunity for thermal sputtering. Therefore it is likely that SN explosions in high density media may explain high redshift objects with such a large amount of dust (e.g. \citealt{2006Hines},  \citealt{2014MNRAS.441.1040R}, \citealt{2015A&A...577A..80M}, \citealt{2016ApJ...816...39M}). 

\section*{Acknowledgements}
We are grateful to our anonymous referee for multiple comments and suggestions which have greatly 
improved the presentation of our results. This study was supported by CONACYT-M\'exico research grant A1-S-28458. SJ acknowledge the support by CONACYT-M\'exico (scholarship registration number 613136) and by the Sistema Nacional de Investigadores (SNI), through its program of research assistants (grant number 620/2018).




\bibliographystyle{mnras}
\bibliography{ref} 



\appendix

\section{The initial configuration of the ejected gas}\label{Ap1}
The ejecta is considered to have a negligible thermal pressure and to be freely expanding. Its velocity then is (\citetalias{1999Mckee}):
\begin{equation}\label{eq:A1}
V\left(R,t\right)=\left\{
	\begin{array}{ll}
		\frac{R}{t}  & \mbox{if }  R \leq R_{ej},\\
		0 & \mbox{if } R > R_{ej},
	\end{array}
\right.
\end{equation}
where $R_{ej}=V_{ej}t$. The initial ejecta mass density is assumed to be:
\begin{equation}\label{eq:A2}
\rho_{ej}\left(R,t \right)=\frac{M_{ej}}{V_{ej}^{3}}f_{n} \left(\frac{V}{V_{ej}} \right)^{-n} t^{-3},
\end{equation}
where $M_{ej}$ and $V_{ej}$ are the ejecta mass and the free-expansion velocity, respectively. $f_{n}$ is a parameter determined by continuity and mass normalization (\citetalias{1999Mckee}). As we are considering cases with $n<3$:
\begin{equation}\label{eq:A3}
f_{n}=\frac{3-n}{4 \pi}, \hspace{0.5cm} n<3.
\end{equation}
The kinetic energy of the ejecta enclosed by the reverse shock, i.e., the energy of the free ejecta, is:
\begin{equation}\label{eq:A3a}
E_{k,free}=2 \pi \int_{0}^{R_{RS}} \rho_{ej}\left(R,t \right) V^{2} R^{2} dR.
\end{equation}
Substituting equation (\ref{eq:A2}) into (\ref{eq:A3a}):
\begin{equation}
E_{k,free}=\frac{1}{2}M_{ej}v_{ej}^{2} \left(\frac{3-n}{5-n} \right)\left(\frac{R_{RS}}{t V_{ej}} \right)^{5-n}.
\end{equation}
The explosion releases a total energy $E_{0}$, which is assumed to be all as kinetic energy of the ejected gas, hence:
\begin{equation}\label{eq:A4}
\frac{E_{0}}{\left(1/2 \right)M_{ej}V_{ej}^{2}}=\frac{3-n}{5-n}, \hspace{0.5cm} n<3.
\end{equation}
The independent parameters are $E_{0}$ and $M_{ej} $. The velocity $V_{ej}$ is calculated from equation (\ref{eq:A4}). \\[0.5cm]
The ejecta mass enclosed by the reverse shock is:
\begin{equation}
M_{free}=4 \pi \int_{0}^{R_{RS}} \rho_{ej}\left(R,t \right)R^{2} dR=M_{ej}\left(\frac{R_{RS}}{V_{ej}t} \right)^{3-n}.
\end{equation}
Therefore, the thermalized ejecta mass is:
\begin{equation}
M_{s2}=M_{ej}\left[1- \left(\frac{R_{RS}}{V_{ej}t} \right)^{3-n}\right].
\end{equation}
This equation can be written in a dimensionless form by making use of equations (\ref{char1}-\ref{char3}) and equation (\ref{eq:A4}):
\begin{equation}\label{Ap:terma}
M^{*}_{s2}=1-\left[2 \left(\frac{5-n}{3-n} \right) \right]^{\frac{3-n}{2} }\left(\frac{R^{*}_{RS}}{t^{*}} \right)^{3-n}. 
\end{equation}
Finally, the momentum carried by the free ejecta is:
\begin{equation}
p_{free}=4 \pi \int_{0}^{R_{RS}} \rho_{ej}\left(R,t \right) V R^{2} dR,
\end{equation}
hence, the total momentum $p_{ej}$ injected by the SN explosion is:
\begin{equation}
p_{ej}=\left( \frac{3-n}{4-n}\right)  M_{ej}V_{ej}.
\end{equation}

\section{The pressure ratio at the beginning of the ED stage}\label{Ap2}
Here, the pressure ratio $\phi$ is estimated at the early ejecta-dominated phase. The pressure gradient between the leading shock $P_{LS}$ and the reverse shock $P_{RS}$ can be estimated from the stationary Euler equation:  
\begin{equation}
\frac{dP}{dr}= -\rho u \frac{du}{dr}.
\end{equation}
At the initial time $t_{0}$: 
\begin{equation}
P_{RS}- P_{LS}\approx -\rho_{LS}U_{LS} \left( U_{RS}-U_{LS}\right),
\label{ap1a}
\end{equation}
where $\rho_{LS}$ is the density behind the leading shock, $U_{LS}$ and $U_{RS}$ are the gas velocities behind the leading and the reverse shock in the rest frame, respectively. The ratio of the gas pressures $\phi$  then is:  
\begin{equation}
\phi=\frac{P_{RS}}{P_{LS}}= 1-\frac{\rho_{LS}U_{LS}}{P_{LS}}\left(U_{RS}-U_{LS} \right).
\label{ap2a}
\end{equation}
From the Rankine-Hugoniot conditions:  
\begin{equation}
P_{LS}=\frac{\gamma+1}{2}\rho_{0} U_{LS}^{2}, \hspace{0.5cm}\rho_{LS}=\frac{\gamma+1}{\gamma-1}\rho_{0},
\label{ap3a}
\end{equation}
where $\rho_{0}$ is the density of the ambient medium. Substituting equation (\ref{ap3a}) into (\ref{ap2a}): 
\begin{equation}
\phi= 1-\frac{2}{\gamma-1}\left(\frac{U_{RS}}{U_{LS}}-1 \right).
\label{ap4a}
\end{equation}
The post-shock velocities are:
\begin{equation}
U_{RS}=\frac{2}{\gamma+1}V_{RS}+\frac{\gamma-1}{\gamma+1}\frac{R_{RS}}{t},
\label{apen1a}
\end{equation}
where $V_{RS}$ and $R_{RS}$ are the velocity and position of the reverse shock. At $t_{0}$:
\begin{equation}
V_{RS}\left(t_{0} \right)=\frac{R_{RS}\left(t_{0} \right)}{t_{0}}=V_{0},
\label{apen2}
\end{equation}
Hence:
\begin{equation}
U_{RS}\left(t_{0}\right)=V_{0}.
\label{ap01}
\end{equation}
The gas velocity behind the leading shock $U_{LS}$ is (see section \ref{initialc}):
\begin{equation}
U_{LS}\left(t_{0} \right)=\frac{2}{\gamma+1}l_{ED}V_{0},
\label{ap02}
\end{equation}
where $l_{ED}=1.1$ is the leading factor. Substituting equations (\ref{ap01}) and (\ref{ap02}) into equation (\ref{ap4a}):
\begin{equation}
\phi\left(t_{0} \right)=1-\frac{2}{\gamma-1}\left(\frac{\gamma+1}{2l_{ED}}-1 \right)=0.3636.
\end{equation}

\section{The initial conditions for the High density test}\label{Ap3}
The ejecta density and its velocity structure presented at \cite{terlevich1992starburst} and related works (e.g, \citealt{1991Franco, 1991Tenorio}) are here discussed. 
\subsection{The density and velocity structure}
The ejected gas is assumed to have a velocity given by:
\begin{equation}
v\left(r,t\right)=\left\{
	\begin{array}{ll}
		\frac{r-R_{c}}{R_{ej}\left( t\right)-R_{c}}  & \mbox{if }  r \geq R_{c},\\
		0 & \mbox{if } r < R_{c},
	\end{array}
\right.
\end{equation}
where:
\begin{equation}
R_{ej}\left( t \right)=R^{0}_{ej}+v_{ej}t,
\label{apend2}
\end{equation}
is the free-expansion radius of the ejecta, $v_{ej}$ its maximum velocity and $R_{c}$ is the inner surface of the ejected mass, i.e., the boundary defining the size of the stellar remnant. The term $R^{0}_{ej}$ is the initial outer boundary of the ejected matter. The mass $M_{ej}$ expelled by the explosion is assumed to be located between $R_{c}$ and $R_{ej}\left(t \right)$:
\begin{equation}
\rho_{ej}\left(r,t\right)=\left\{
	\begin{array}{ll}
		\frac{M_{ej}}{4 \pi \ln \left(R_{ej}\left(t \right) /R_{c}\right)}r^{-3}   & \mbox{if }  r \geq R_{c},\\
		0 & \mbox{if } r < R_{c},
	\end{array}
\right.
\label{apend3}
\end{equation}
The fraction of thermalized ejecta mass is now given by:
\begin{equation}
M_{th}=M_{ej}(1-\frac{\ln\left(R_{RS}/R_{c} \right)}{\ln\left(R_{ej}\left(t \right)/R_{c} \right)})
\label{apend5}
\end{equation}
The parameters $R_{ej}^{0}$ and $R_{c}$ are calculated from the initial conditions fulfilling the data from \cite{terlevich1992starburst}. Indeed, The authors set $M_{ej}=2.5$ M$_{\odot}$, and state that an initial energy $E=10^{51}$ erg and momentum $p_{0}= 2.44 \times 10^{42}$ g cm s$^{-1}$ were deposited into an ambient medium of $n_{0}=10^{7}$ cm$^{-3}$. Table \ref{table1} presents the set of parameters that satisfy these initial conditions.  

\begin{table}\label{table1}
\caption{Initial conditions for a SNR evolving into a medium of density $n_{0}=10^{7}$ cm$^{-3}$.}
\begin{center}
\begin{tabular}{c c }
\hline
$M_{ej}[$M$_{\odot}]$ & 2.5 \\
$v_{ej}[$km s$^{-1}]$&$1.4 \times 10^{4}$\\
$E_{0}$ [erg] & $10^{51}$\\
$R_{ej}^{0}[10^{-2}$ pc] & 0.70\\
$R_{c}[10^{-2}$ pc]& 0.11\\
\hline
\end{tabular}
\end{center}
\end{table}

\bsp	
\label{lastpage}
\end{document}